# Enhanced Thermal Transport through Soft Glassy Nano-disc Paste

Susheel S. Bhandari and K. Muralidhar

Department of Mechanical Engineering

Indian Institute of Technology Kanpur

Kanpur 208016 INDIA

Yogesh M Joshi*

Department of Chemical Engineering

Indian Institute of Technology Kanpur

Kanpur 208016 INDIA

* Corresponding author, Email: joshi@iitk.ac.in

**Abstract**

We study diffusion of heat in an aqueous suspension of disc shaped nanoparticles of Laponite, which has finite elasticity and paste-like consistency, by using the Mach-Zehnder interferometer. We estimate the thermal diffusivity of the suspension by comparing the experimentally obtained temperature distribution to that with analytical solution. We observe that despite highly constrained Brownian diffusivity of particles owing to its soft glassy nature, suspensions at very small concentrations of Laponite demonstrates significant enhancement in thermal diffusivity. We correlate the observed enhancement with the possible microstructures of the Laponite suspension.

PACS: 44.10.+I, 44.35.+c, 65.60.+a

## I. Introduction

With advancing technologies and continuously shrinking devices such as electronic gadgets, chemical (micro) reactors, and fuel cells, there is an ever increasing demand for a medium with enhanced thermal transport properties. Heat can, in principle, be transferred by three mechanisms: conduction, convection and radiation.[1] However, at miniature scales, small gaps, and at moderate temperature gradients, conduction is the dominant mode of heat transfer. An important requirement for effective heat transfer to the environment in media such as heat sink pads is close contact between the sink and the instrument without gaps or air pockets. It necessitates the material to be easily deformable and strong enough to support the weight of the gadget. In this work we study heat transport characteristics of a soft glassy solid with paste-like consistency. It comprises an aqueous suspension of nano-disks of Laponite, a model soft glassy material.[2] We believe that this system is unique due to oblate (disk-like) anisotropic shape of the suspended particles, whose heat transport behavior has scarcely been studied in the literature. This soft solid like suspension is observed to demonstrate remarkable enhancement in thermal diffusivity at very low volume fractions.

In order to enhance performance, a single phase material is routinely mixed with a small percentage of additive or filler having significantly different physical properties. If both the major and the minor components are in the solid state, the resultant combination is represented as composite. If the major component is liquid and minor component is solid particles, we get a suspension which retains the viscous properties of the liquid in the limit of small fraction of solid particles. In both situations, if the physical dimension of the solid particles is in nanometric range, the former is called nanocomposite, while recent literature describes the latter as nanofluid. With regard to heat transfer behavior, suspensions including nanofluids can transport heat by conduction as well as convection mechanism. However, owing to their fluid character, they do not possess enough physical strength necessary to support heat producing gadgets. Composites, on the other hand, can transport heat only by conduction. In addition composite do possess the necessary strength to support weight, but cannot provide a very close contact without gaps.

Suspensions of particles having size in the micrometer range are observed to follow the Maxwell relation that relates the effective thermal conductivity to its volume fraction ($\phi$). This relation suggests enhancement to scale as $3\phi$, when particle thermal conductivity is much larger than the suspending liquid conductivity.[3-5] However, the recent literature suggests that, compared to micron sized particles, suspensions of nanoparticles, namely, nanofluids, show significant enhancement in thermal conductivity when compared at equal volume fractions.[6, 7] Typically, an enhancement of 10 to 120 % for particle volume fraction of 2 % is reported in the literature for various nanofluids,[3, 4, 6] whereas the Maxwell relation predicts a maximum enhancement of only 6 %. Various mechanisms proposed to explain this unusual behavior include phonon transport,[6] Brownian diffusion,[8] molecular ordering on nanoparticles,[9] and cluster/aggregate formation.[10-12] However, most of these mechanisms cannot completely explain the anomalous increase in thermal conductivity available in the reported experimental data. On the other hand, there is broad consensus in recent literature on the possible role of aggregate formation in the observed increase in thermal conductivity.[3, 6]

Heat transport of both the micro-composites as well as nanocomposites has been studied in great detail.[13] Typically the conductivity of the suspended medium, extent of aggregation, its aspect ratio and orientation, quality of interface between suspended particles, and the density of percolation in the matrix determines heat transport behavior.[14-18] However, contrary to that observed in nanofluids, some reports suggest that thermal conductivity of nanocomposites is lower than that seen in micro-composites for a given loading of the filler.[16] Such a phenomenon is attributed to the enhanced resistance associated with the particle – matrix interface, and is expected to have pronounced effect at nanoscopic dimensions.

In this work we study heat transport behavior of an aqueous suspension of disc-like nanoparticles of Laponite RD which has soft-solid like consistency, and therefore is intermediate to nanofluids and nanocomposites. Laponite RD is a synthetic silicate clay mineral having a disc-like shape with diameter of 25 nm and thickness of 1 nm.[19] Although not available in the literature, thermal conductivity of a Laponite particle is expected to be higher than water owing to its structure as a single crystal.[20] In an aqueous medium of pH 10, the Laponite disk is known to possess negative charge on opposite faces with a weakly positive charge on its edge. Addition of salt screens the negative charge thereby enhancing attraction among Laponite particles. Overall, a combination of attractive and repulsive electrostatic forces leads to complex interaction among the particles,[21-24] so that incorporation of Laponite in water increases its viscosity and elasticity by over several orders of magnitude eventually rendering a soft solid-like consistency.[2, 25, 26] Laponite suspension is also known to possess yield stress.[27] Several scattering studies have shown that such massive increase in viscosity and elasticity is due to caging of particles causing strong reduction in Brownian diffusivity of disc-shaped nanoparticles.[28, 29] Consequently, Laponite suspension undergoes ergodicity breaking and owing to its pasty nature it is represented as a soft glass.[29]

**II. Experimental Procedure and Material Preparation**

In the present work we estimate thermal diffusivity of soft glassy aqueous Laponite suspension using laser interferometry through an unsteady heat diffusion experiment. In a typical experiment as shown by a schematic in figure 1, a slab of material of thickness $H$ and thermal diffusivity $\alpha$ is maintained at temperature $T_C$. At time $t=0$, the temperature of the top plate is suddenly raised to $T_H$. The thermal field follows a diffusion equation: $\partial \theta / \partial \tau = \partial^2 \theta / \partial \eta^2$, where $\theta = (T - T_C)/(T_H - T_C)$, $\tau = \alpha t / H^2$ and $\eta = y/H$.[30] For a constant $\alpha$, this diffusion equation can be analytically solved for known initial and boundary conditions ($\theta = 0$ at $\eta = 0$; $\tau \leq 0$, $\theta = 0$ $\forall$ $\eta$; $\tau > 0$, $\theta = 1$ at $\eta = 1$) to yield the time-evolution of temperature field as:

$$\theta = \eta - 2\sum_{n}\left[(-1)^{n+1}/n\pi\right]\sin(n\pi\eta)\exp(-n^2\pi^2\tau), \quad n=1, 2, 3... \qquad (1)$$

By comparing experimental results with eq. (1), parameter $\alpha$ can be estimated.

We procured Laponite RD® from Southern Clay Products Inc. The suspension of Laponite in ultrapure water is highly transparent. Suspension was prepared in water having pH 10 and predetermined amount of salt (NaCl) using the procedure described elsewhere.[23] A freshly prepared suspension was stored in a sealed polypropylene bottle for around 30 to 50 days. Before carrying out the interferometry experiment, an aged suspension was vigorously stirred. Presheared Laponite suspension was poured in the test chamber and left undisturbed for around four hours at constant temperature $T_C$. In this work, we have used the Mach-Zehnder interferometer, wherein the split collimated laser beam [632.8 nm, 35 mW He-Ne laser (spectra-physics)] is passed through the test and the reference chambers. The two beams are superimposed, before being imaged by a CCD camera as shown in a schematic figure 2. The optical path difference created during the experiment between the reference and the test beams produces an interference pattern. The reference chamber was filled with glucose solution. Before beginning the experiment, refractive index of glucose solution was matched with the sample in the test section ensuring no optical path difference between the two light beams. Subsequently, the temperature of the top plate was increased to $T_H$. Temperatures adopted in various experiments are $T_C$ = 24 to 28°C and $T_H$ = 27 to 30°C (the bottom plate temperature maintained using a constant temperature bath is usually kept equal to the ambient temperature which varied between 24 to 28°C.). The hot and cold surfaces were held at spatially and temporally uniform temperatures by constant temperature baths. As time passes, heat diffuses from the heated top plate towards the bottom increasing temperature within the medium. Change in temperature alters refractive index of the suspension creating an optical path difference between the test and the reference chambers, thereby producing an interference pattern.

Aqueous suspension of Laponite is known to demonstrate thixotropy, leading to time and deformation field dependent viscosity and elasticity [2, 27].

Under such conditions, elastic and viscous modulus gives a reliable measure of the rheological behavior of the sample. We measure elastic and viscous modulus of 30 days old sample. In a typical experiment we apply an oscillatory shear stress ($\sigma = \sigma_0 e^{i\omega t}$) having magnitude $\sigma_0$ =0.5 Pa and frequency $\omega$ =0.1 Hz to the samples 4 hours after shear melting the same in a rheometer. The experiments were carried out at 25°C using MCR 501 rheometer. Owing to applied stress field, strain induced in the material has a form: $\gamma = \gamma_0 e^{i(\omega t - \delta)}$. The elastic ($G'$) and viscous ($G''$) moduli are given by: $G' + iG'' = \sigma_0 e^{i\delta}/\gamma_0$.[31] The heat capacity of dry Laponite powder was also measured using differential scanning calorimeter and was found to be 1.03 kJ/(kgK).

**III. Results and Discussion**

Subsequent to an increase in temperature of the top plate as a step function, heat diffusion towards the lower plate gradually increases the suspension temperature, leading to an optical path difference between the test and reference chambers. In figure 3, we describe the corresponding evolution of the interference pattern. Due to diffusion of heat from the top surface, fringes appear in its vicinity and migrate towards the lower surface with time. In the limit of large time, steady state sets in and fringes get uniformly distributed. The temperature difference between two consecutive fringes ($\Delta T_\varepsilon$) can be estimated from the principles of wave optics and is given by: $\Delta T_\varepsilon = \lambda/(L dn/dT)$, where $\lambda$ is wavelength of light, $L$ is length of test cell, and $dn/dT$ is variation of refractive index with temperature.[32] We used values of $dn/dT$ for Laponite suspension available in the literature.[33] Therefore, with the knowledge of temperature at one boundary and location of fringes, the complete thermal field can be obtained.[34] In figure 4, we plot evolution of temperature as a function of time from the interference pattern shown in figure 3. These results are essentially equivalent to the analytical solution [eq. (1)] for various time instants. At short times the temperature data is not very reliable due to sharp gradients that cause refraction of the light beam. In addition, in the limit of short and large times, the temperature field is not sensitive to the value of $\alpha$. Closer to steady state, the temperature field in the suspension is affected by the thermal properties of the confining surfaces. Quantitatively, sensitivity of the estimated parameter to the measurement is expressed by: $S(t,\alpha) = \int_0^H (dT/d\alpha) dy$.[35] In order to estimate $\alpha$, we fit eq. (1) to the experimental data obtained in that window where $S(t,\alpha)$ is high by a procedure that is equivalent to the method of weighted least squares. In figure 4 the fit of the analytical solution (displayed by lines) is shown to only that data which is contained in the 90 % sensitivity window. It can be seen that the data fits the analytical solution very well. It should be noted that the height of the test cell is $H$ =50 mm and the sample is stably stratified in density. Hence, there are no convection currents, and the present method measures the bulk thermal diffusivity $\alpha$ of the solution.

In figure 5, we plot $\alpha$ as a function of Laponite concentration ($\phi_L$). For pure water ($\phi_L$ =0), the analysis of interferometry data estimates the value of $\alpha$ very close to that given in the literature ($\approx 1.4\times 10^{-7}$ m$^2$/s). It can be seen that incorporation of Laponite in water leads to considerable enhancement in $\alpha$. In the inset of figure 5, we plot effect of concentration of NaCl ($C_s$) on $\alpha$ for 0.8 and 1 % Laponite concentration suspensions. It can be seen that, within the experimental uncertainty, $\alpha$ changes weakly with change in $C_s$. We also estimate thermal conductivity ($k=\rho C_P \alpha$) of Laponite suspension from the experimentally determined heat diffusivity data (figure 5). Heat capacity of the aqueous suspension is obtained by using the rule of mixtures $C_{P,mix} = \sum_i w_i C_{Pi}$, where $w_i$ and $C_{Pi}$ are mass fraction and heat capacity of $i$ th fraction respectively. In figure 6, $k$ for the Laponite suspension is plotted as a function of $\phi_L$, while in the inset we plot the same as a function of $C_s$. It can be seen that the data follows the same trend as thermal diffusivity shown in figure 5.

We also carried out oscillatory rheological experiments on the suspension samples. As mentioned in the introduction, aqueous suspension of Laponite for concentration range (of Laponite as well as salt) explored in this work has soft solid like consistency. In figure 7 we plot elastic and viscous modulus of aqueous suspension of Laponite as a function of concentration of Laponite. Irrespective of the concentration of Laponite in the explored range, we observe elastic modulus to be always greater than viscous modulus ($G' >> G''$). It has been observed that elastic modulus of aqueous suspension of Laponite is a very weak function of frequency over practically explorable frequency range and therefore should be termed as a soft solid (with infinite viscosity).[36] Such rheological behavior also indicates significantly constrained translational diffusivity of the Laponite particles in the suspension. Similar to the sample without salt, samples having salt also shows qualitatively similar rheological behavior (not shown in figure) with a soft solid-like nature.[27] The results of figures 5, 6 and 7 are striking as significant enhancement in $\alpha$ is observed despite constrained Brownian diffusivity.[28, 29]

The microstructure of Laponite suspension which restricts Brownian diffusivity remains a matter of discussion.[21, 22, 36, 37] Some groups have proposed that aqueous suspension of Laponite typically above 0.8 - 1 volume % (2 – 2.5 weight %), forms a repulsive glass.[21, 24] In this state the Laponite particles, owing to electrostatic repulsion among them, remain in self suspended state (without touching each other). On the other hand, some groups believe aqueous suspension of Laponite is a gel formed by an interconnected network of fractal aggregates of particles,[22] which form a percolation path. The precise microstructure responsible for enhanced thermal transport in the present system, whether repulsive glass or an attractive gel, is a subject of continuing discussion.[36, 38] We postulate the following reason for the enhancement of heat transport in the Laponite suspension, viewed from independent perspectives: as glass as well as gel.

In case Laponite suspension forms a repulsive glass wherein particles do not touch each other, heat transport can occur primarily via transfer of thermal motion. Typically for 0.8, 1 and 1.2 volume % Laponite suspensions, average inter-particle distances are equal to 40, 37 and 34 nm respectively, the average inter-particle distance decreasing as $\sim \sqrt[3]{\phi_L}$.[21] However, owing to the anisotropic shape of the particle (a disk with 25 to 30 nm diameter and 1 nm thickness), the nearest points can be significantly closer than the average value. Laponite particle, due to its crystalline nature, is expected to have higher thermal conductivity than water. Therefore, transfer of heat can get enhanced through the particles in the suspension. Decrease in distance between the particles caused by increase in concentration is anticipated to enhance heat transport further as shown in figures 5 and 6.

When Laponite suspension is visualized to form a network of fractal aggregates, heat transport can be seen as occurring on two scales: one, the aggregates that fill the space and the other, within the aggregate where Laponite discs form a fractal network. In this proposal Laponite particles touch each other in an edge-to-face fashion. Hence, transfer of heat can take place more effectively through the percolation network. In this case, the volume fraction of aggregates $\phi_a$, which is also indicative of the density of percolation network, is related to $\phi_L$ as $\phi_L = \phi_i \phi_a$,[11] where $\phi_i$ is volume fraction of Laponite particles within an aggregate. Increase in $\phi_i$, enhances thermal conductivity of the aggregate, though at the expense of density of the percolation network. Therefore, this picture requires a fine balance between $\phi_i$ and $\phi_a$, so as to cause enhancement in overall heat transport. According to Nicolai and coworkers,[22] density of a three dimensional network increases with Laponite concentration, thereby making available more paths per unit area for the heat transport.

As mentioned before, nanofluids are also observed to demonstrate significant enhancement in thermal conductivity compared to the base fluid. The recent literature on nanofluids suggests the important role played by fractal aggregates in enhancing its thermal conductivity.[3, 6] It should be noted that, the rheological constitution of nanofluids with respect to that of the present system is completely different. Nanofluids are essentially in liquid state, have finite viscosity and no measurable elasticity. Aqueous suspension of Laponite, on the other hand, has infinite viscosity (in the limit of weak deformation fields), finite elasticity with pasty or soft solid-like consistency. Additionally, owing to formation of aggregates, sedimentation may cause nanofluids to become instable.[39, 40] On the other hand sedimentation is not observed in Laponite suspension beyond the concentration of 0.4 volume %.[24]

The present study is unique from various points-of-view. First, very little work is available in the literature, experimental as well as theoretical, regarding heat transport through suspension of oblate (disc-like) particles. The present system has a consistency of a paste and is described in the recent literature as soft glass.[2] Very little work has been done on such systems to study their heat transport characteristics. Additionally, interferometry as a technique, which determines properties of materials in bulk, is being used for the first time to estimate heat transport parameters in suspensions. In the present study we report remarkably enhanced bulk heat transport in a paste-like medium at very small concentrations of the suspended medium. We believe that the results mentioned in this paper will attract new applications for pasty materials and stimulate further discussions.

**IV. Conclusion**

In this work we investigate conduction heat transfer in pasty suspension of nano-disks of Laponite in water using a Mach-Zehnder interferometer. Elastic modulus of Laponite suspension is observed to be significantly greater than viscous modulus indicating the material to be in solid state with highly constrained Brownian diffusivity of particles in the suspension. In the heat transfer experiments transient evolution of temperature as a function of space is obtained using interferometry. The thermal diffusivity of the suspension is obtained by comparing experimentally obtained time dependent temperature profile with the analytical expression. We observe that thermal diffusivity increases with concentration of Laponite. Remarkably the material shows significant enhancement for very small concentrations of Laponite.

**Acknowledgement:** This work was supported by Department of Science and Technology, Government of India. We thank Mr. Yogesh Nimdeo for measuring heat capacity.

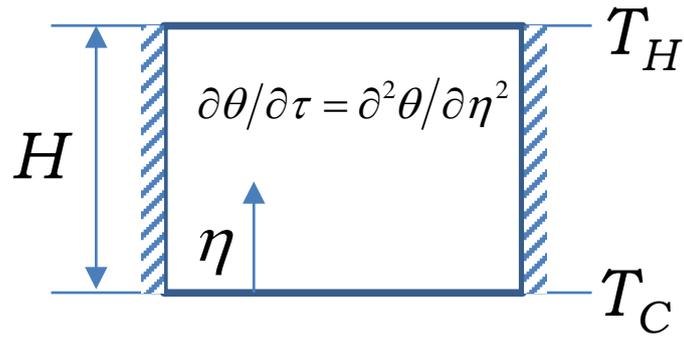

**Figure 1.** (Color online) Representation of thermal field. Slab of Laponite suspension having thickness $H$ is maintained at temperature $T_C$ for $t \leq 0$. For $t > 0$, temperature of top surface is increased to $T_H$.

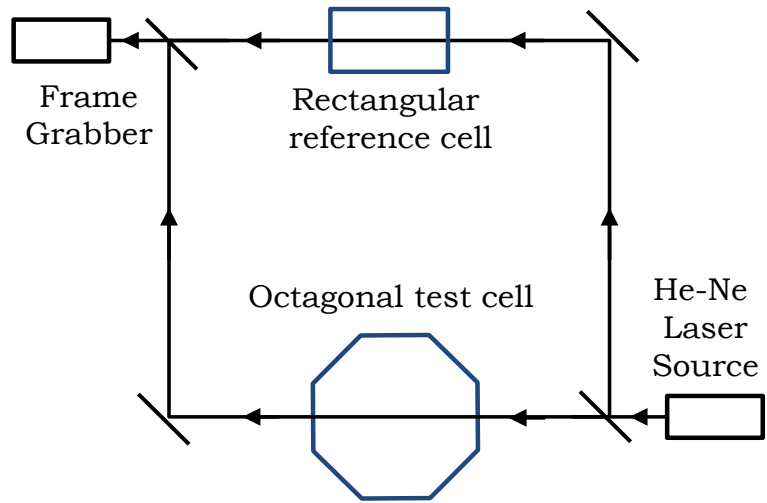

**Figure 2.** (Color online) Schematic of Mach-Zehnder interferometer. The test cell has an octagonal cross section. This cell is a top view of arrangement shown in figure 1. The test beam and reference beam meet to produce an interference pattern depending upon the path difference between the two.

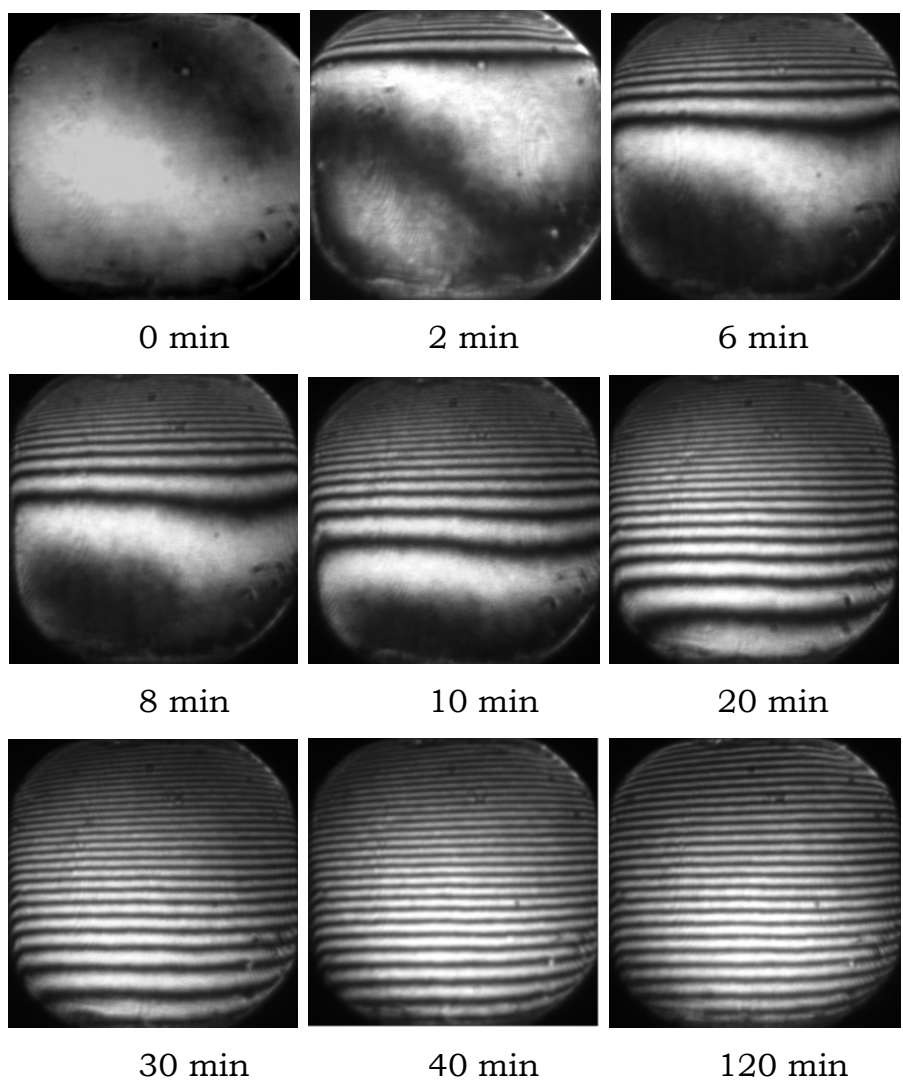

**Figure 3:** Evolution of interference pattern as a function of time for 0.8 volume % aqueous Laponite suspension. The upper and lower surfaces were maintained at 27°C and 25°C respectively.

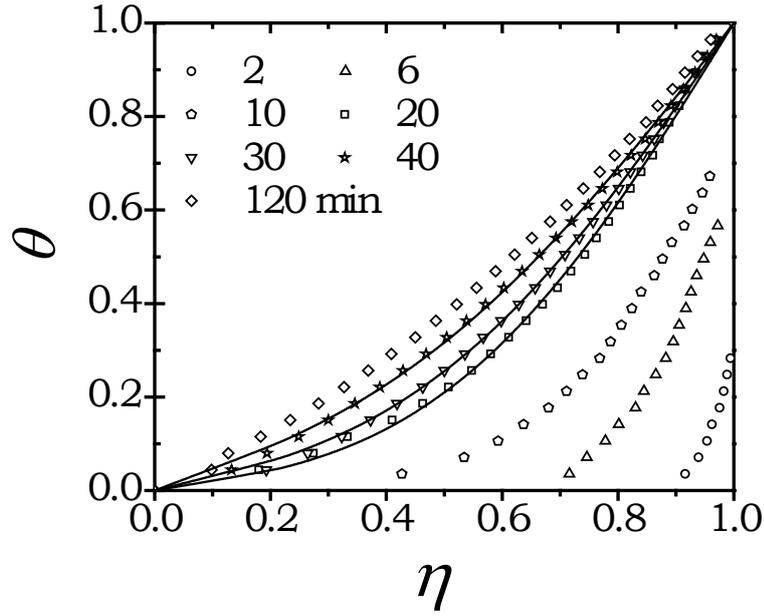

**Figure 4:** Evolution of dimensionless temperature as a function of dimensionless distance from the bottom plate at various times for interference patterns shown in figure 3. Symbols are experimental data while lines are fits of eq. (1) to the experimental data.

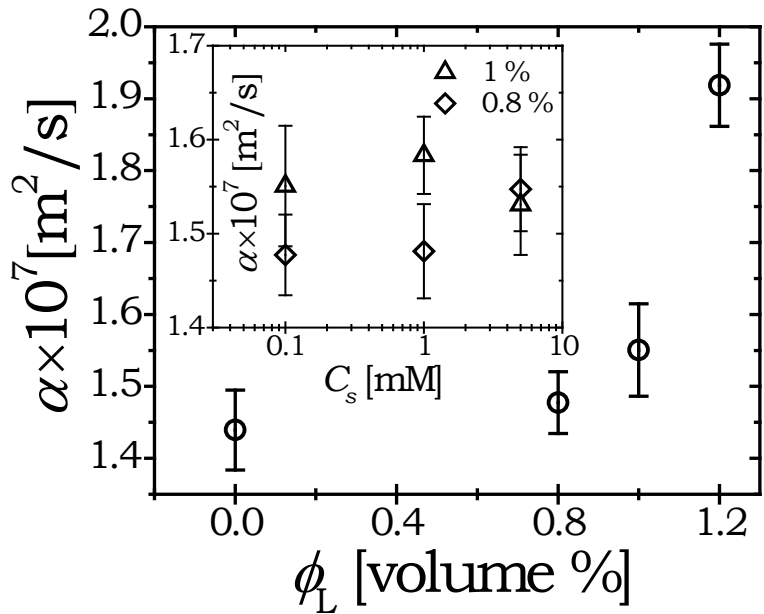

**Figure 5:** Thermal diffusivity of aqueous Laponite suspension is plotted as a function of volume % of Laponite ($\phi_L$) without any externally added salt ($C_S$ =0.1 mM). The inset shows the effect of concentration of Na$^+$ ions ($C_S$) on thermal diffusivity of aqueous suspension of Laponite for $\phi_L$=0.8 and 1 volume %.

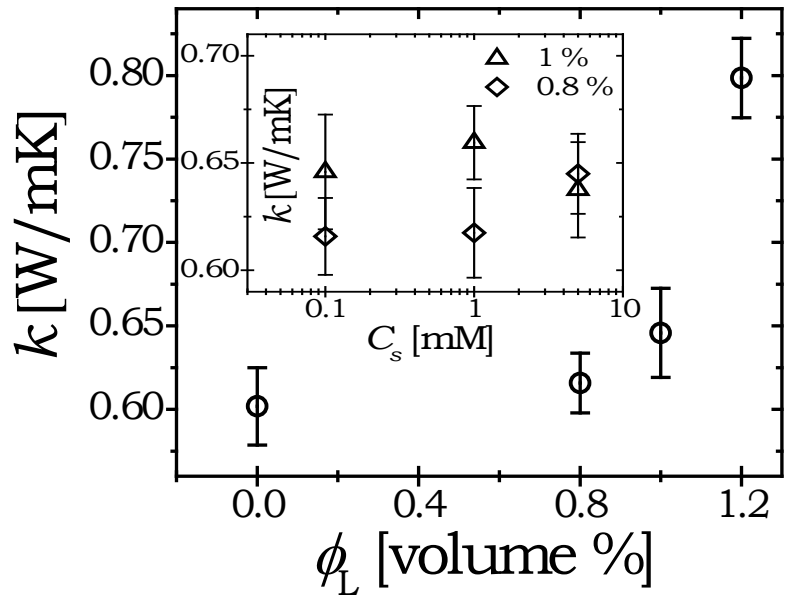

**Figure 6:** Thermal conductivity of aqueous Laponite suspension is plotted as a function of volume % of Laponite ($\phi_L$) for the data shown in figure 5. The inset shows the effect $C_S$ on thermal conductivity of aqueous suspension of Laponite for $\phi_L$=0.8 and 1 volume % for the data shown in figure 5.

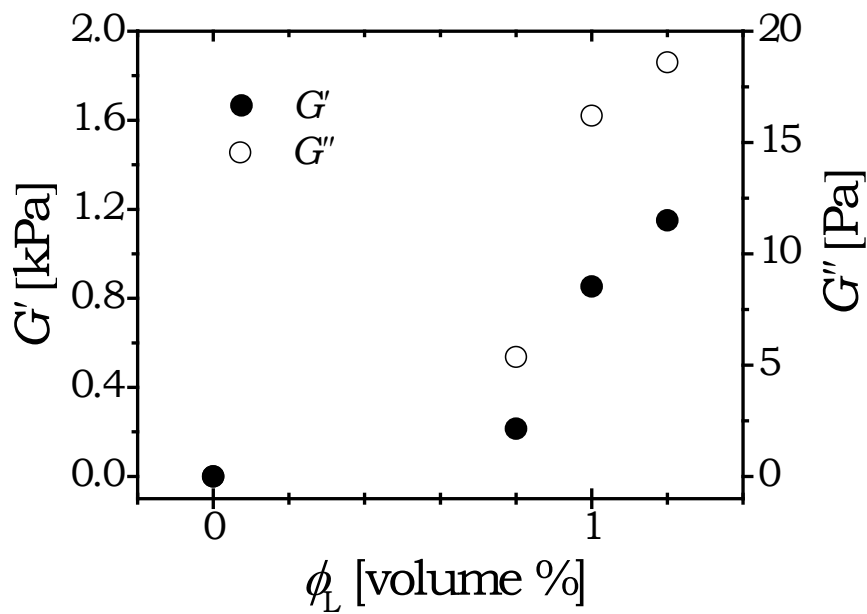

**Figure 7.** Elastic modulus (filled circles) and viscous modulus (open circles) of aqueous Laponite suspension is plotted as a function of volume % of Laponite ($\phi_L$).